%% file: main.tex
\documentclass[modern, twocolumn]{aastex62}
\usepackage{amsmath}
\graphicspath{{./}{figures/}}

\received{}
\revised{}
\accepted{}
\submitjournal{ApJ}

\shorttitle{Multimessenger gamma-ray and neutrino coincidence alerts}
\shortauthors{Ayala Solares et al.}

\begin{document}

\title{Multimessenger Gamma-Ray and Neutrino Coincidence Alerts using HAWC and IceCube subthreshold Data}

\correspondingauthor{Hugo A. Ayala Solares}
\email{hgayala@psu.edu}


\input{authors_AMON}
\input{authors_HAWC}
\input{authors_IC}








\begin{abstract}

The High Altitude Water Cherenkov (HAWC) and IceCube observatories, through the Astrophysical Multimessenger Observatory Network (AMON) framework, have developed a multimessenger joint search  for extragalactic astrophysical sources. This analysis looks for sources that emit both cosmic neutrinos and gamma rays that are produced in photohadronic or hadronic interactions.
The AMON system is running continuously, receiving subthreshold data (i.e. data that are not suited on their own to do astrophysical searches) from HAWC and IceCube, and combining them in real time. 
Here we present here the analysis algorithm, as well as results from archival data collected between 2015 June and 2018 August, with a total live time of 3.0 years. During this period we found two coincident events that have a false-alarm rate (FAR) of $<1$ coincidence yr$^{-1}$, consistent with the background expectations. The real-time implementation of the analysis in the AMON system began on 2019 November 20 and issues alerts to the community through the Gamma-ray Coordinates Network with an FAR threshold of $<4$ coincidences yr$^{-1}$. 

\end{abstract}



\section{Introduction} \label{sec:intro}

The coincident detection of gravitational waves and electromagnetic radiation~\citep{bns}, as well as the evidence found for a neutrino coincident with a gamma-ray flare from the blazar TXS 0506+056~\citep{icneutrino}, have shown the potential of multimessenger astrophysics. 
The ability to combine data from different observatories in real time or near-real time is driving this new era in astrophysics. 
The Astrophysical Multimessenger Observatory Network (AMON) has been created to facilitate the interaction of different observatories, create a framework for analyses with distinct datasets across multiple experiments, and notify the astrophysical community of any interesting events worthy of follow-up~\citep{amonI,amonII}\footnote{AMON website:~\url{https://www.amon.psu.edu/}}.

AMON focuses on using data that are below the discovery threshold of individual observatories. These events by themselves are heavily background-dominated, which complicates a search for astrophysical sources. By statistically combining the temporal and/or spatial information of these subthreshold events provided by different detectors, AMON aims to recover the signal events that are hidden among the background of each single observatory.
Two multimessenger analyses were previously developed combining gamma-ray data from Fermi-LAT with neutrino data: one analysis using IceCube data~\citep{Turley:2018} and the other using ANTARES data~\citep{Turley:2019}.\footnote{Although what constitutes the data depends on the groups or collaborations, in general, the position and time of the events are always used. Other information is added if available.} The Fermi-LAT and ANTARES coincidence search started running in real time in 2019 April and has issued two alerts to date \citep[see GCN circulars][]{FA1,FA2}.

In this work, we focus on a new coincidence analysis combining information from the High Altitude Water Cherenkov (HAWC) Gamma-Ray Observatory~\citep{hawc} and the IceCube Neutrino Observatory~\citep{icecube} using the AMON infrastructure. This new multimessenger channel has been operational as a real-time coincidence search since 2019 December. 

The purpose of this analysis is to search for hadronic accelerators that produce both gamma rays and neutrinos as secondary particles, with an emphasis on transient events.  The accelerated cosmic rays can interact with target material surrounding the environment of the sources or with radiation fields. These interactions produce charged and neutral pions. Charged pions predominantly decay via $\pi^{+} \rightarrow \mu^{+} + \nu_{\mu}$, followed by the decay of the muon as $\mu^{+} \rightarrow e^{+} + \nu_e + \bar{\nu}_{\mu}$ (and charge conjugate). Neutral pions decay into two gamma-ray photons, $\pi^{0} \rightarrow \gamma + \gamma$. The ratio between charged pions and neutral pions depends on the type of interaction of the cosmic rays with the targets. If the interaction occurs with electromagnetic radiation, the interaction will be photohadronic, which produces charged and neutral pions with probabilities of one-third and two-thirds, after considering both resonant and nonresonant pion productions. If the pions originate from interactions of cosmic rays with matter, the probability of producing charged and neutral pions is one-third for each type of pion~\citep{neutrinoProd}. A useful relation between the fluxes of gamma rays ($F_{\gamma}$) and neutrinos ($F_{\nu_{\alpha}}$) is expressed as 
\begin{equation}\label{eq:nuem}
    E_{\gamma} F_{\gamma}(E_{\gamma}) \approx e^{-\frac{d}{\lambda_{\gamma \gamma}}} \frac{2}{3K}\sum_{\nu_{\alpha}} E_{\nu} F_{\nu_{\alpha}} (E_{\nu}),
\end{equation}
where $E_{\gamma}\approx 2E_{\nu}$ are the gamma ray and neutrino energies; $\alpha$ corresponds to the neutrino flavor; $K$ is the ratio of charged to neutral pions, with $K=1$ for photohadronic interactions and $K=2$ for hadronuclear interactions; $d$ is the distance to the source; and $\lambda_{\gamma \gamma}$ accounts for the attenuation of gamma rays due to their interaction with the extragalactic background light (EBL)~\citep[see][]{prdkohta2,prdkohta}.


In this paper, we present the algorithm and analysis to search for possible sources of gamma rays and neutrinos by looking at HAWC's and IceCube's subthreshold data. In section \ref{sec:data}, we describe briefly the detectors and their data. In section \ref{sec:method}, we present the statistical method and provide the false-alarm rate (FAR), sensitivities and discovery potentials. 

In section \ref{sec:results}, we present the results obtained using 3 years of archival data, including upper limits for the same period of time for the total isotropic equivalent energy and source rate density parameter space. Finally, we conclude and discuss the implementation of the analysis in real-time using the AMON framework. 

\section{HAWC and IceCube detectors and datasets} \label{sec:data}

HAWC and IceCube are two detectors that focus on high-energy astrophysics, searching for sources that accelerate cosmic rays. Both detectors use the Cherenkov technique where photomultipliers are used to detect the Cherenkov light produced by the passage of secondary charged relativistic particles---from gamma ray, neutrinos, and cosmic-ray showers---through a medium. HAWC uses water as the medium, while IceCube uses the Antarctic ice. 

Due to the attenuation of gamma rays on the extragalactic background photons, the signal from a source might not be significantly detected above background in the HAWC data. However, if IceCube neutrino events are found in spatiotemporal coincidence with a subthreshold HAWC hot spot, this might become an interesting coincidence that could be followed up by other observatories. In addition, the uncertainty region of HAWC events is generally smaller compared to IceCube events, which can give a better localization of a potential joint source. 

\subsection{High-energy Gamma Rays from HAWC}
The HAWC observatory is a high-energy gamma-ray detector located in central Mexico. 
The complete detector has been in operation since 2015 March. 
HAWC has a large field of view, covering two-thirds of the sky every day with a high-duty cycle in the declination range from $-26^{\circ}$ to $64^{\circ}$. HAWC is mainly sensitive to gamma rays in the energy range between $300$~GeV and $100$~TeV. It has an angular resolution of $0.2^{\circ}{-}1.0^{\circ}$ (68\% containment) that depends on the energy of the event, its zenith angle and size of the shower footprint measured by HAWC \citep{hawc}.

\begin{figure}
    \centering
    \includegraphics[scale=0.26]{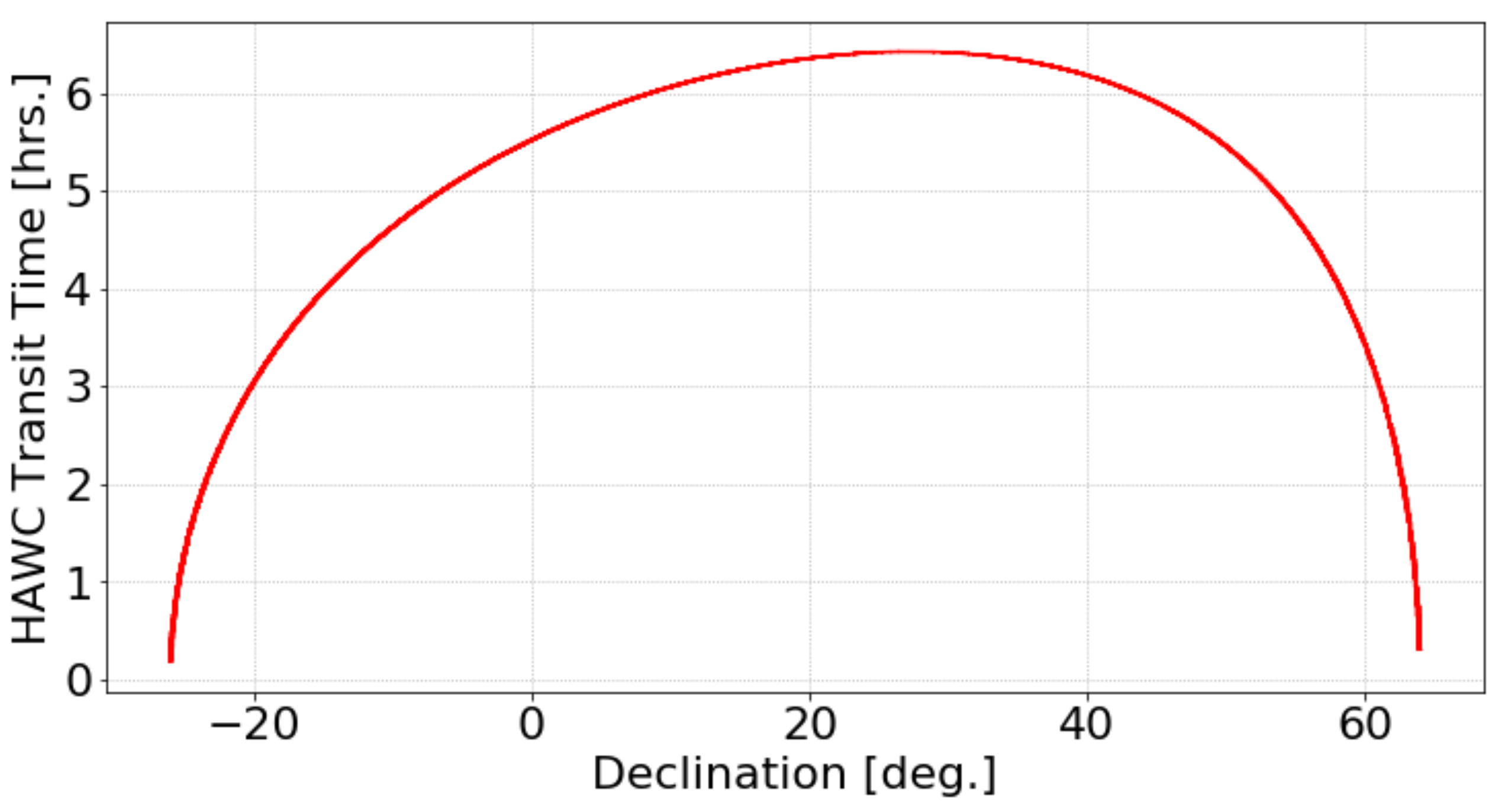}
    \caption{Duration of a transit of a point in the sky as a function of declination above the HAWC detector, applying a zenith angle cut of $<45^{\circ}$. }
    \label{fig:transittime}
\end{figure}

We select locations of excess exceeding a given significance threshold--- called ``hot spots'' ---from the HAWC data to be used as inputs to the combined search. Hot spots are defined as locations in the sky with a cluster of events above the estimated cosmic-ray background level and measured by the significance (excess above the background). They are identified during one full transit of that sky location above the detector.
The main hot-spot parameters AMON receives are: the position coordinates and their uncertainty; significance value, with a minimum of $2.75\sigma$ (threshold defined by HAWC); and the start and stop times of the transit. The duration of the transits are declination dependent as shown in Fig. \ref{fig:transittime}. 
Since we are searching for unknown sources or sources that cannot be significantly detected above the background, we mask the data from the following parts of the sky above HAWC: the Galactic plane ($b<|3^{\circ}|$), the Crab Nebula, Geminga, Monogem, Mkr 421 and Mkr 501. The current rate of these hot spots received by AMON is $\sim$800 per day.

\subsection{High-energy Neutrinos from IceCube}
The IceCube observatory is  a detector of high-energy neutrinos located at the South Pole~\citep{icecube}.  It became fully operational in 2011 after 7 years of construction. IceCube first observed the high-energy astrophysical neutrino flux in 2013 \citep{iceastro}.

IceCube can search for neutrinos from the whole sky, though it is more sensitive to sources from the northern celestial hemisphere, since the Earth helps reduce the atmospheric background in IceCube. This is an advantage in this analysis since HAWC is primarily sensitive in the northern sky. 
IceCube is sensitive to energies that can reach up to 1 EeV near the horizon (declination of $0^{\circ}$).
The angular resolution depends on the topology of the events inside the detector. Two main topologies are observed: track events and cascade events. Track events are mostly induced by charged-current muon-neutrino interactions. These tracks can have a length of several kilometers and most of the time extend beyond the detector volume. The track events have a median angular resolution of ${\sim}0.4^{\circ}$ above 100 TeV. 
Cascade events are produced by the other types of neutrinos or neutral-current interactions of any neutrino type. They have better energy resolution compared to tracks, since the energy deposited by the events is completely contained inside the detector. Their angular resolution, however, is ${>}10^{\circ}$ with current reconstruction methods \citep{icecubegfu}.


The IceCube candidate events sent to AMON consist of single throughgoing tracks. These events can have energies above 0.1~TeV for upgoing events, while downgoing events can have energies above 100~TeV. Lower-energy events are more probable to be background events. The parameters consist of the sky position and its uncertainty, the time of the event, and the reconstructed energy or boosted decision tree (BDT)\footnote{The BDT score is used to reduce the atmospheric muon background as well as separate the astrophysical signal.} score \citep[see Section 3 of][]{icecubegfu}, depending on whether the event is in the northern or southern hemisphere, respectively, and can be used to calculate the background $p$-value of the event. The current rate of the events received by AMON is $\sim$650 per day. 

\section{Method} \label{sec:method}

The coincidence analysis is applied to events satisfying two criteria. The first is a temporal selection requiring the neutrino events to arrive within the transit time of the HAWC hot spot. Second, we select neutrinos that are within a radius of $3.5^{\circ}$ from the HAWC hot-spot localization.\footnote{The angular distance is motivated from IceCube multiplet searches \citep[see][]{multiplet}.} 
After the neutrino events have passed the selection criteria, we calculate a statistic to rank the coincident events. The rate of coincidences after passing the criteria is $\sim$100 per day.
This ranking statistic is based on Fisher's method~\citep{fisher}, where we combine all the information that we have from the events. It is defined as

\begin{equation}\label{eq:rank1}
\chi^2_{6+2n_{\nu}}=-2 \ln[p_{_{\lambda}}p_{_{\rm HAWC}}p_{_{\rm cluster}}\prod_i^{n_{\nu}}p_{_{\rm IC,i}}],
\end{equation}
where the number of degrees of freedom is ${6+2n_{\nu}}$ (as described below). The quantity  $p_{_{\lambda}}$ quantifies the overlap of the spatial uncertainties of the events. The value $p_{_{\rm HAWC}}$ is the probability of the HAWC event being compatible with a background fluctuation. Since we can expect more than one IceCube candidate event in the time window (i.e. the HAWC transit period), we can calculate the probability of background IceCube events occurring in that time window. Given that we have at least one event detected, the $p_{_{\rm cluster}}$\footnote{Here $p_{_\mathrm{cluster}(n_{\nu})} = 1 - \sum^{n_{\nu}-2}_{i=0} \mathrm{Pois}(i;f_{\nu} \Delta T $), where $f_{\nu}$ is the IceCube background rate and $\Delta T$ is the HAWC transit time.} is the probability of that one event to be in the same time-window with the observed number of IceCube events, $n_{\nu}$, or more from background; if there is only one IceCube event, this value is equal to 1.0. The value $p_{_{\rm IC,i}}$ is the probability of measuring a similar or higher energy/BDT score for an IceCube event, assuming it is a background event (calculated using the energy/BDT score and zenith angle). 
The $p_{_{\lambda}}$ value is obtained by a maximum-likelihood method that measures how much the positions of the HAWC and IceCube events overlap. This is calculated as 
\begin{equation}\label{eq:lambda}
\lambda(\boldsymbol{x}) = \sum_{i=1}^N  \ln(\frac{S_i(\boldsymbol{x})}{B_i}),
\end{equation}
where $N$ is the HAWC hot spot plus the number of IceCube candidate events. $S$ corresponds to a signal directional probability distribution function, which is assumed to be a Gaussian distribution on the sphere with a width given by the measured positional uncertainty from each detector, $S_i(\boldsymbol{x}) =  \exp{[-(\boldsymbol{x}-\boldsymbol{x}_{i})^2/2\sigma_{i}^2}]/(2 \pi \sigma_{i}^2)$. $B_i$ is the background directional probability distribution from each detector at the position of the events. 
The position of the coincidence, $\boldsymbol{x}_\mathrm{coinc}$, is defined as the position of the maximum likelihood value, $\lambda_\mathrm{max}$, as shown in Figure \ref{fig:skymaps}.  
The uncertainty of $\boldsymbol{x}_\mathrm{coinc}$ is calculated by the standard error $\sigma_{\boldsymbol{x}_\mathrm{coinc}}^2=1/\sum_i^N(\sigma_i^{-2})$. 

The $\lambda_\mathrm{max}$ values are used to make a distribution of the overlap of the coincidences. A higher $\lambda_\mathrm{max}$ value indicates a more significant overlap of the event uncertainties. This translates into a smaller $p$-value $p_{_{\lambda}}$.

Due to the fact that we can have more than one IceCube event passing the selection criteria, the degrees of freedom of Eq. \ref{eq:rank1} vary. We therefore calculate a $p$-value of the $\chi^2$ with ${6+2n_{\nu}}$ degrees of freedom. The ranking statistic (RS) is then simply defined as $-\log_{10}(\text{$p$-value})$. 

\subsection{Calibration of the FAR}\label{sec:simultaion}

\begin{figure*}
    \centering
    \includegraphics[scale=0.5]{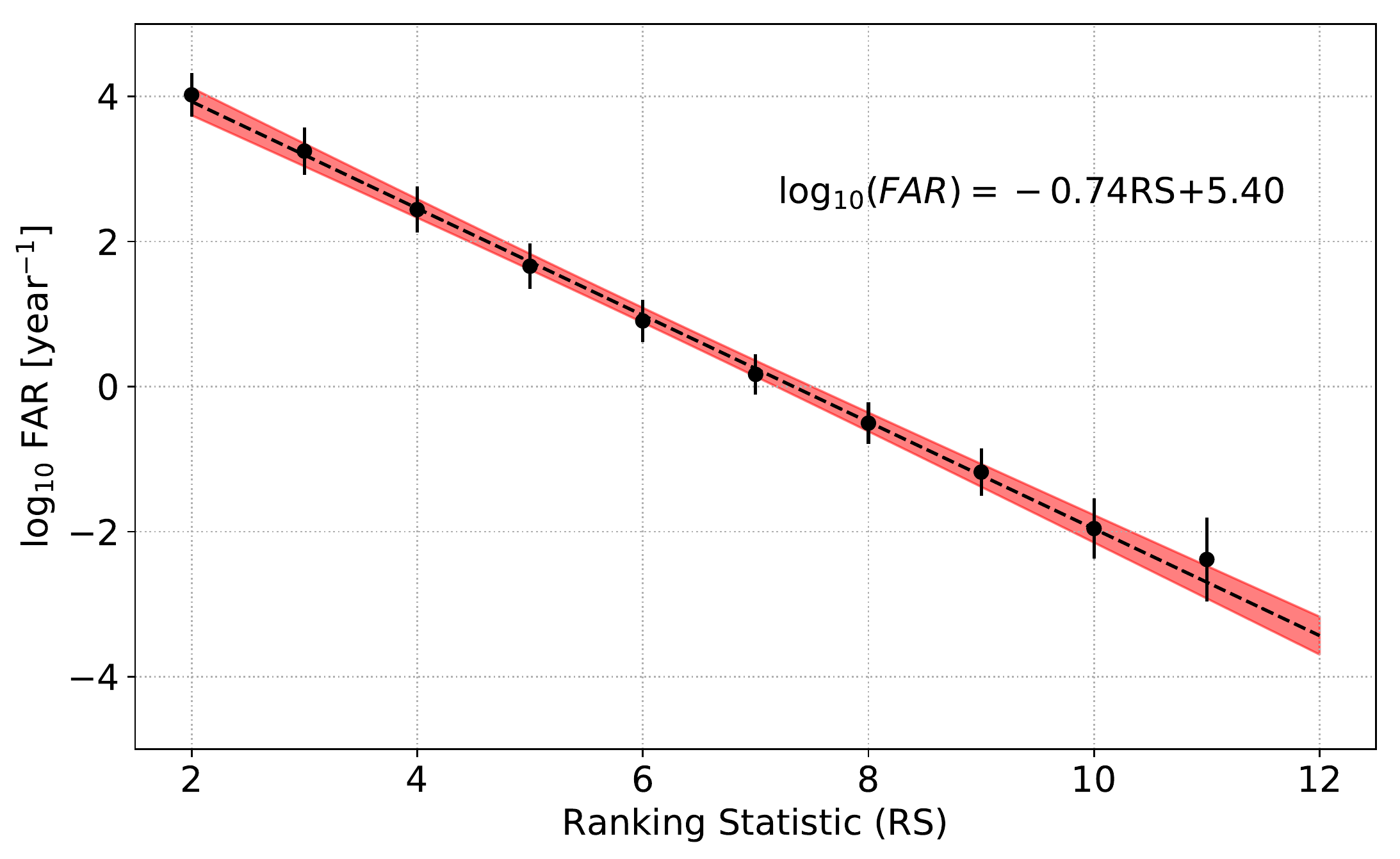}
    \caption{The FAR as a function of the ranking statistic obtained from the scrambled data sets. The width of the band (in red) is the 1$\sigma$ statistical uncertainty. The function in the graph will be used to select alerts that will be sent to the Galactic Coordinates Network~\citep{gcn}. A false-alarm rate of 1 per year is obtained with a ranking statistic value of 7.3.}
    \label{fig:rates}
\end{figure*}

We apply the above-described algorithm to 3 years of scrambled data sets from both observatories. 
Scrambling consists of randomizing the right ascension and time values of the events many times in order to calibrate the FAR. 
The result of this process is shown in Fig. \ref{fig:rates}. For a specific ranking statistic, we calculate the total number of coincidences above this ranking statistic value and then divide by the total amount of scrambled simulation time to get the rate. 
The linear fit in Fig. \ref{fig:rates} is used to estimate the FAR in real-time analyses. 

\subsection{Sensitivity and Discovery Potential} \label{sec:sensitivity}
To put the archival results into context, we look at a simulation for transient events that can produce both neutrinos and gamma rays.
We quantify the sensitivity and discovery potential for the 1 coincidence per year threshold for a live time of 3 years of data.

\begin{figure*}[!htp]
    \centering
    \includegraphics[scale=0.4,clip]{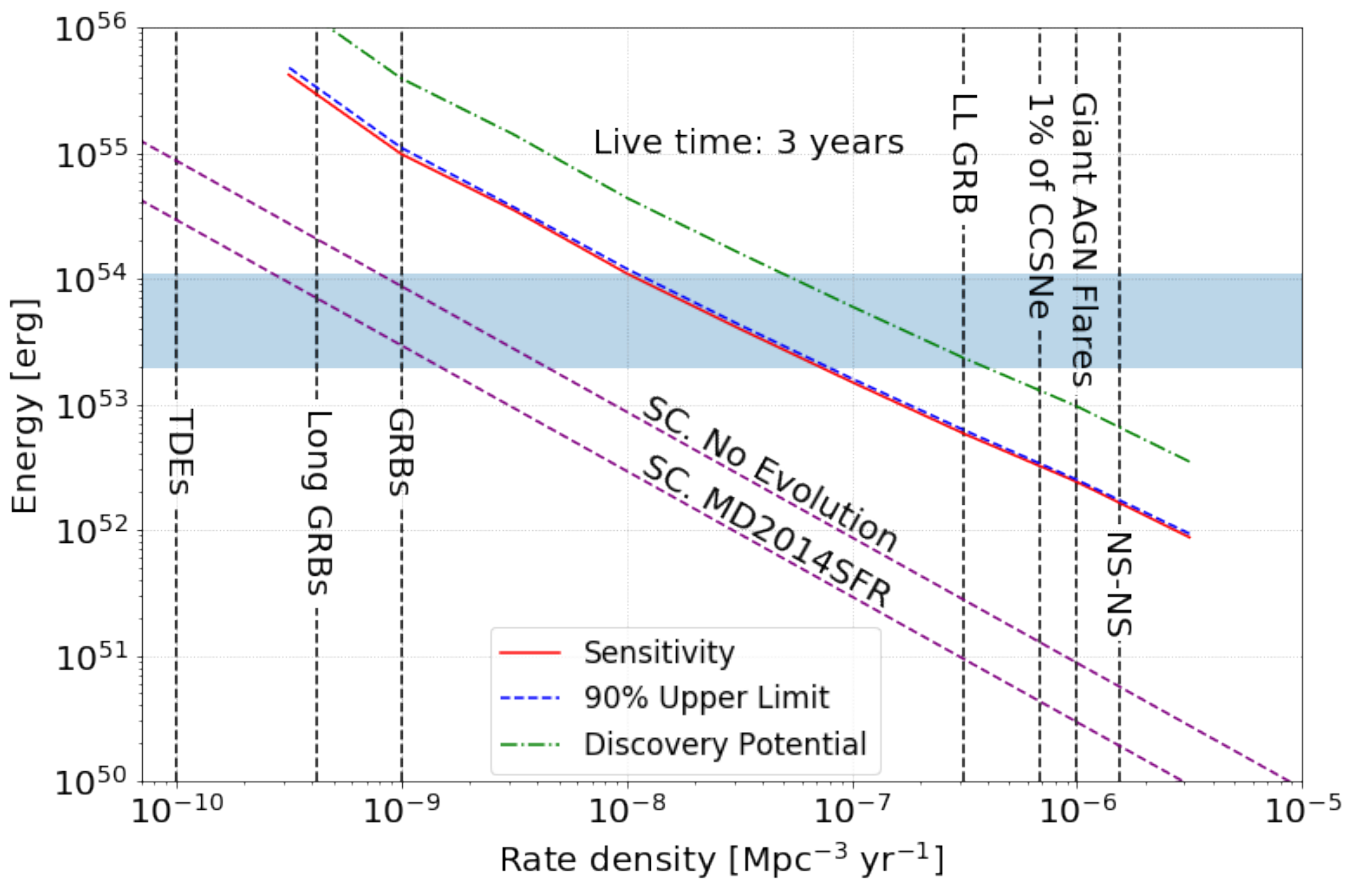}
    \caption{Sensitivity (red), discovery potential (green) for the 3 years of data as a function of rate density and total isotropic equivalent energy in neutrinos of transients of the order of 6 hours and assuming a power-law spectrum with index of -2.0. The number of coincidences below the 1 per year FAR threshold is used as the statistic. The upper limit (blue) result is explained in Section \ref{sec:upperlimit}. The results are higher by a factor of 3 if a power-law spectrum with index -2.4 is assumed (See text for more details). The light-blue horizontal band corresponds to the estimated released neutrino energy of the event IceCube-170922A related to TXS 0506+056 \citep{icneutrino}. The purple lines are the total neutrino isotropic equivalent energy of the source as a function of rate density that would be required to produce the total observed IceCube neutrino diffuse emission with neutrino energies between 100 GeV and 10 PeV\citep{prdkhohta3}. The vertical lines correspond to different source rate densities \citep{ccsn,grb_rate_II,gaf,sourcesI}. The comparison is valid under the assumption that the transient phenomena are of the order of hours.  
    \label{fig:sensUL}}
\end{figure*}

We use the FIRESONG software package \citep{firesong}, which simulates neutrino sources for a given local rate density of transient gamma-ray and neutrino sources, total neutrino isotropic equivalent energies, and timescales. The outcome of the simulation is a list of simulated neutrino sources with declination, redshift and neutrino flux normalization. This is based on a power-law energy spectrum with spectral index of -2 for the flux, in the energy range between 10 TeV and 10 PeV,\footnote{The simulation was also run with a spectral index of -2.4. Since the energy range for IceCube's sensitivity changes with index, the range was extended from 100 GeV to 10 PeV. The sensitivity and discovery potential of the analysis are higher by a factor of 3. Figure \ref{fig:sensUL} shows the result for the simulation with a spectral index of -2.0.} and a time of the burst of 6 hrs.\footnote{Since the information given by HAWC is averaged over one transit, we use this timescale for the simulations.} 
Using Eq. \ref{eq:nuem}, we can transform the normalization to a gamma-ray flux assuming photohadronic interactions. We then simulate the sources in HAWC, adding EBL attenuation with the model from \cite{eblDominguez} and in addition, we draw a Poisson random number of neutrinos with an expectation value given by the source flux and IceCube's background. Finally, we quantify the coincidence.

We calculate the sensitivity and discovery potential by running simulations for a given pair of rate density and total neutrino isotropic energy. We apply the coincidence algorithm and after finding the signal coincidences, they are added to a distribution with random coincidences. We keep the total number of coincidences the same as that of the 3 years of data, so we remove the same number of random coincidences as injected sources. We apply this procedure several times in order to build a distribution of the number of coincidences that cross the 1 coincidence per year threshold, $N(FAR\leq1)$. 
If no sources are injected, $N(FAR\leq1)$ is a Poisson distribution with a rate of $r_B = 3.0$ (B stands for background) for the 3 years of observations. 
For the sensitivity, we find the pair of parameters that will give us a $r_B + r_S = 6.0$ (where S stands for signal). This corresponds to a $N(FAR\leq1)$ distribution that crosses the median of the Poisson background distribution 90\% of the time.
For the $5\sigma$ discovery potential, we find the pair of parameters that will give a rate of $r_B + r_S = 15.7$ since this distribution will have 50\% of its population with a $p$-value smaller than $2.87\times10^{-7}$ with respect to the Poisson background distribution.
We fit the distribution of $N(FAR\leq1)$ to a Poisson function and find the best value for $r_S$. The pair of rate density and total neutrino isotropic energy that gives the corresponding $r_S$ values for sensitivity or discovery potential is plotted in Fig. \ref{fig:sensUL}.
To put the sensitivity and discovery potential in context, we include diagonal lines that show the total neutrino isotropic energy as a function of rate density that would be required to produce the total observed IceCube diffuse neutrino flux (assuming a power-law spectrum with index of -2.5). 
This assumes either no evolution or the star-formation evolution following the Madau-Dickinson model \citep{MD2014SFR}; it also assumes a standard candle (SC.) luminosity function. Based on \cite{icneutrino}, we marked a region on Fig. \ref{fig:sensUL} showing the estimated released neutrino energy of the IceCube event 170922A related to TXS 0506+056.

\section{Results}\label{sec:results}

\subsection{Archival Data}
\begin{figure*}
    \centering
    \includegraphics[scale=0.5]{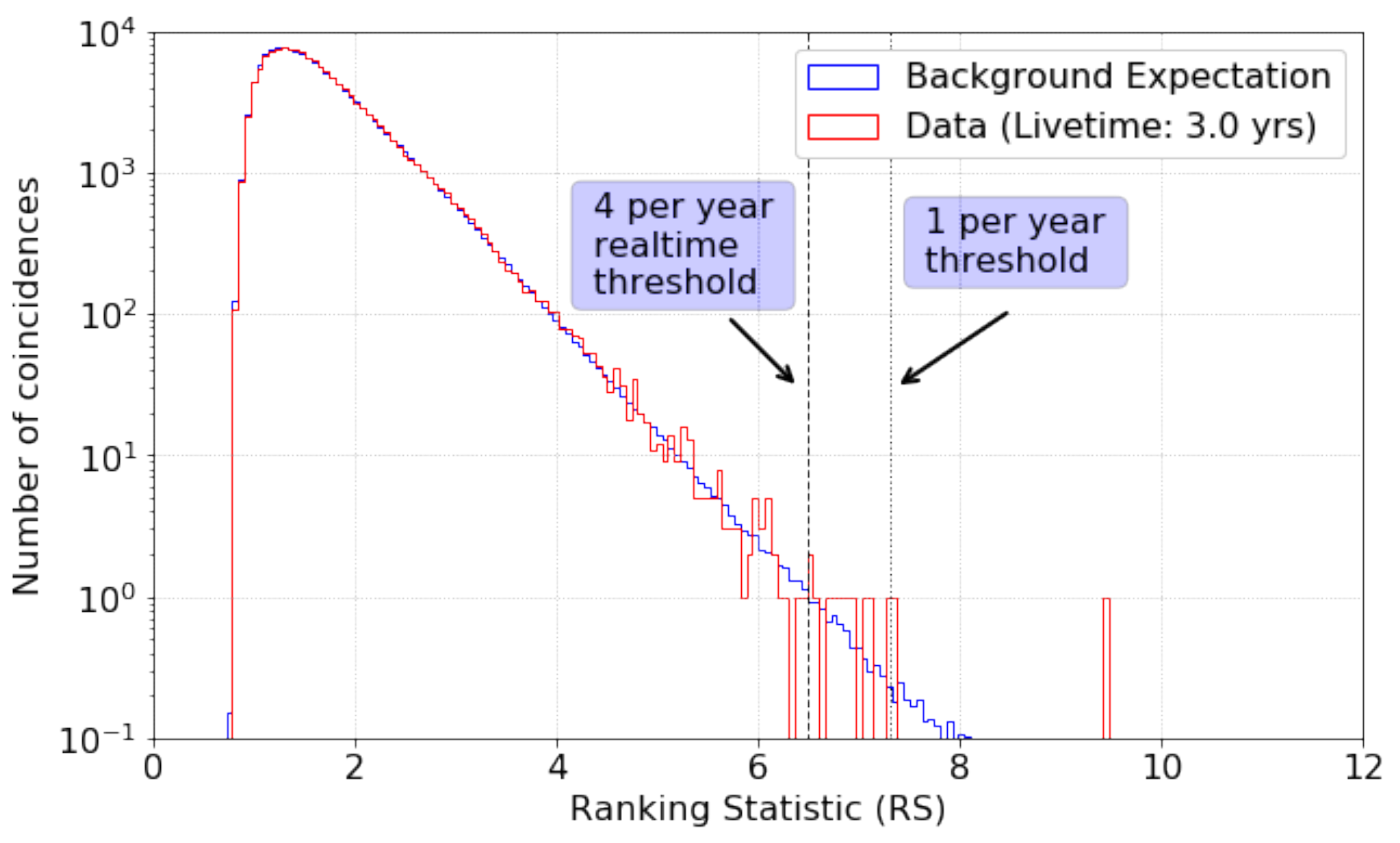}
    \caption{Ranking statistic distribution of the analysis.  Blue:  background expectation obtained from the scrambled data sets and normalized to the number of coincidences observed in the unblinded data set.  Red: result from the unblinded analysis.  Live-time is 3 years of data. The vertical lines mark 4 and 1 coincidence per year thresholds. The highest ranking statistic in the 3 year dataset is 9.4 (1 every 38.5 years).}
    \label{fig:coincdist}
\end{figure*}

We analyzed data collected from June 2015 to August 2018.
Fig. \ref{fig:coincdist} shows the distribution of ranking statistic value of the unblinded data compared to the expected distribution of random coincidences (i.e. scrambled datasets mentioned in Section \ref{sec:method}). 

\begin{figure*}%
	\centering
	 \includegraphics[scale=0.5,clip]{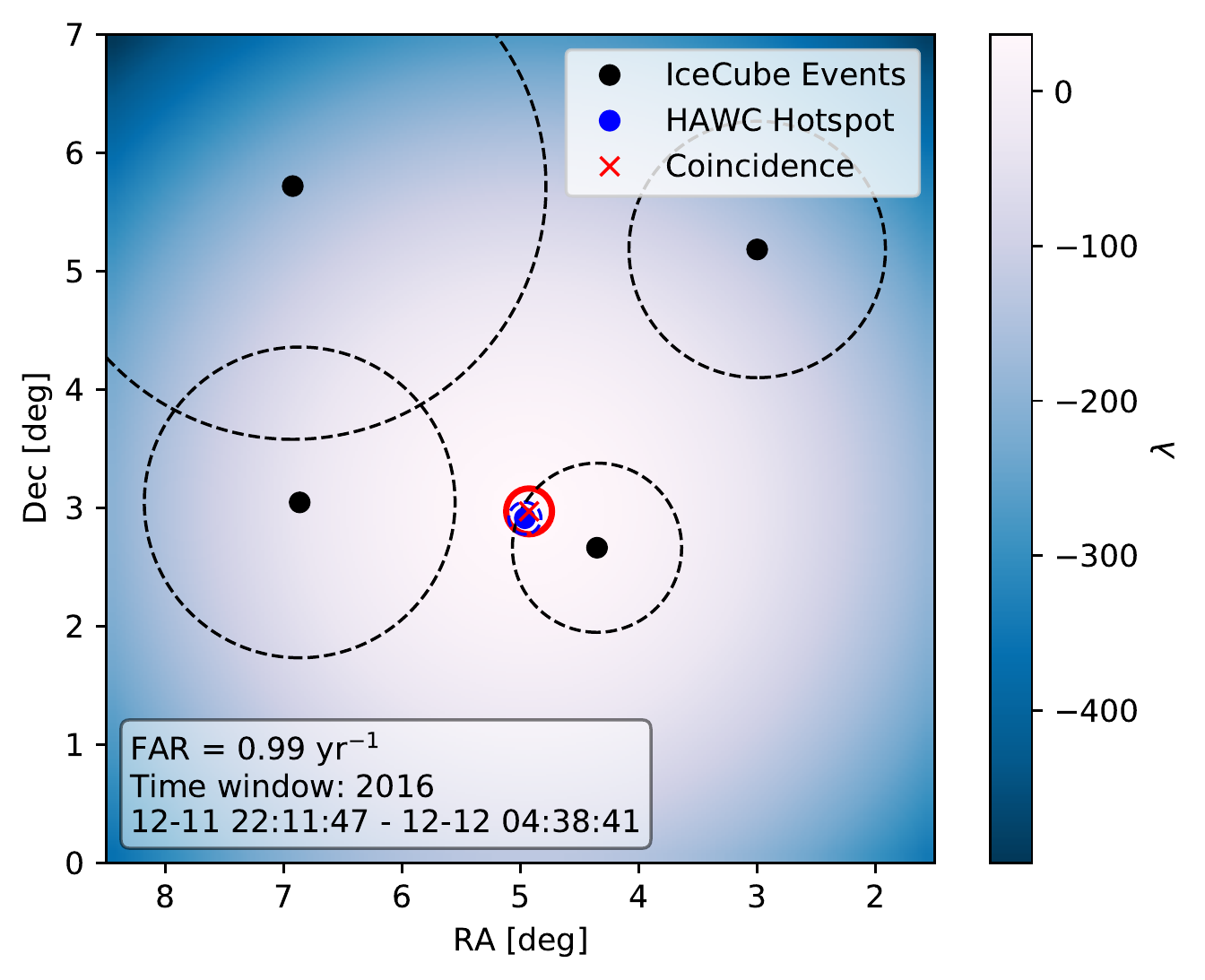}
	 \includegraphics[scale=0.5,clip]{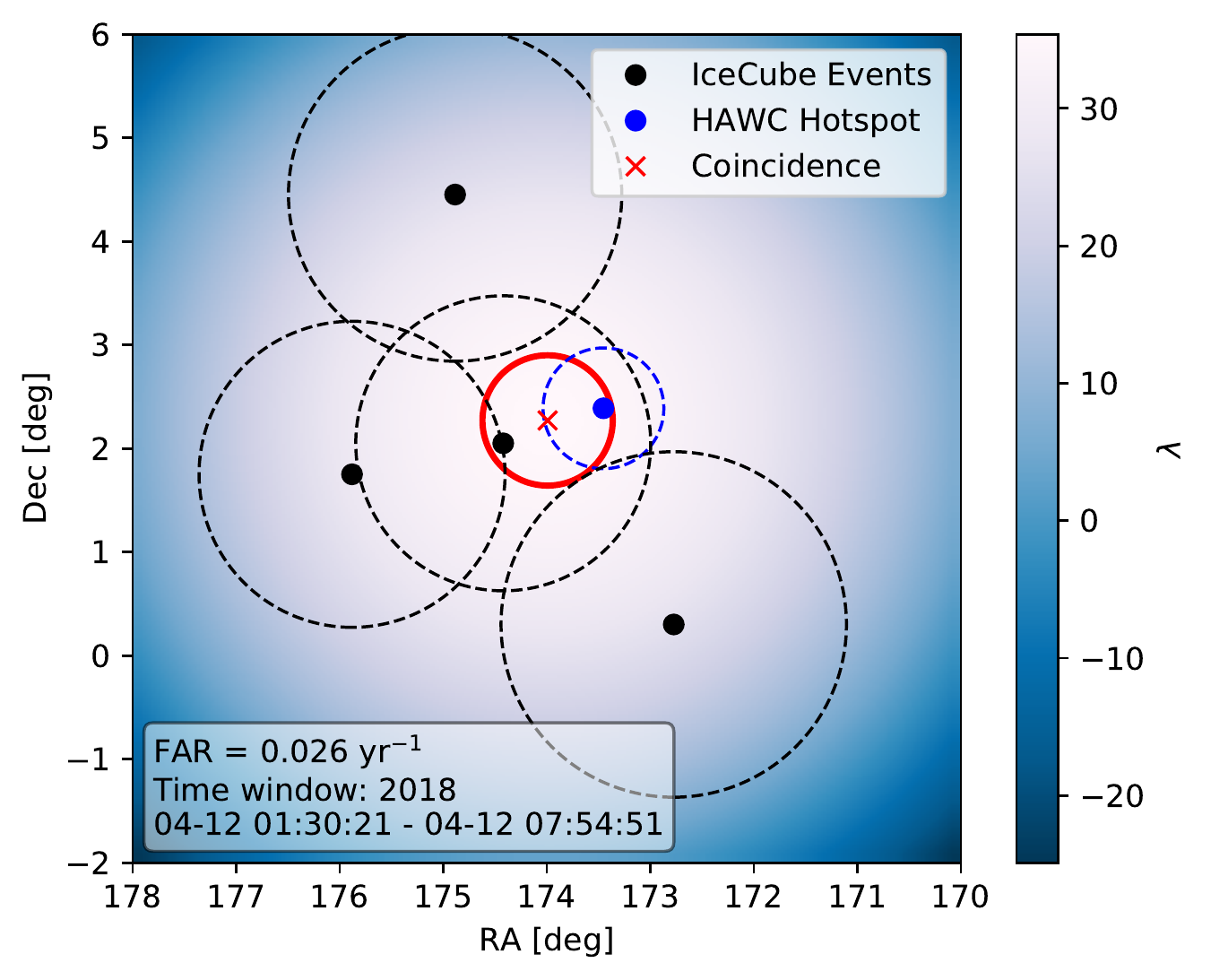}
	\caption{Skymaps of the coincidences with the lowest FAR found in the 3 years of archival data. Position of the individual events are marked with the dots. The best-fit combined positions $\boldsymbol{x}_\mathrm{coinc}$,  found after optimizing Eq. \ref{eq:lambda}, are marked with a cross. Circles are the 50\% containment region. }
	\label{fig:skymaps}
\end{figure*}

Since we are interested in searching for rare coincidences, we look for  coincidences with an FAR of less than 1 coincidence per year, which corresponds  to a ranking statistic value of 7.31. 
We found two coincidences, one in 2016 and one in 2018, with ranking statistics of 7.34 (1 coincidence per year) and 9.43 (1 coincidence in 38.5 years) respectively. These coincidences are not significant with respect to the background distribution. Using $\text{$p$-value} = 1-\exp{(-t\cdot{FAR})}$, with $t=3$ years, the $p$-values are 0.95 and 0.075 respectively. The skymaps of the two coincident events with the highest ranking statistic values are shown in Fig.~\ref{fig:skymaps}. Table~\ref{tab:Coincs} contains the summary information on them. Information of the individual events that form each coincidence can be found in Tables \ref{tab:CoincHAWC} and \ref{tab:CoincIC}. 

\begin{deluxetable*}{cccccc}
\tablecaption{Summary information on the two coincidences with FAR $<$ 1\label{tab:Coincs}}
\tabletypesize{\scriptsize}
\tablehead{
\colhead{Dec [deg]} & \colhead{RA [deg]} & \colhead{Uncertainty (50\% containment)[deg]} & \colhead{Ranking Statistic} & \colhead{FAR [per year]} & \colhead{$p$-value} 
}
\startdata
2.96 & 4.93 & 0.16 & 7.3 & 0.99 & 0.95 \\
\hline
\hline
2.27 & 173.99 & 0.53 & 9.4 & 0.026 & 0.075\\
\enddata
\end{deluxetable*}


\begin{deluxetable*}{ccccccc}
\tablecaption{Information on the two HAWC ``hot spots'' that correspond to each of the coincidences with an FAR$<1$ per year in the 3 year Data set\label{tab:CoincHAWC}. }
\tabletypesize{\scriptsize}
\tablehead{
\colhead{Dec} & \colhead{RA} & \colhead{Uncertainty} & \colhead{Initial Time} & \colhead{Final Time} & \colhead{Significance} & \colhead{Flux upper limit } \\
\colhead{[deg]} & \colhead{[deg]} & \colhead{[deg]} & \colhead{[UT]} & \colhead{[UT]} & $\sigma$ & \colhead{[TeV$^{-1}$ cm$^{-2}$ s$^{-1}$]}
}
\startdata
2.91 & 4.96 & 0.17 & 2016-12-11 22:11:47 & 2016-12-12 04:38:41 & 3.71 & 3.9e-11\\
\hline
\hline
2.38 & 173.4 & 0.74 & 2018-04-12 01:31:21 & 2018-04-12 07:54:51 & 2.77 & 8.3e-11\\
\enddata
\tablenotetext{}{\textbf{Note.} Flux upper limits are based on a $E^{-2}$ energy spectrum.}
\end{deluxetable*}

\begin{deluxetable*}{cccccc}
\tablecaption{IceCube neutrino information for each of the coincidences\label{tab:CoincIC}.}
\tabletypesize{\scriptsize}
\tablehead{
\colhead{Dec} & \colhead{RA} & \colhead{Uncertainty} & \colhead{Time} & \colhead{Background $p$-value} & \colhead{$\Delta \theta$} \\
\colhead{[deg]} & \colhead{[deg]} & \colhead{[deg]} & \colhead{[UT]} & \colhead{p$_{_{IC}}$ }& \colhead{[deg]} 
}
\startdata
3.04 & 6.86 & 1.31 & 2016-12-11 23:20:25 & 0.944 & 1.90\\
2.66 & 4.35 & 0.71 & 2016-12-12 00:24:48 & 0.055 & 0.65\\
5.18 & 3.00 & 1.08 & 2016-12-12 01:37:28 & 0.391 & 2.99\\
5.71 & 6.92 & 2.13 & 2016-12-12 03:22:12 & 0.993 & 3.42\\
\hline
\hline
0.30 & 172.77 & 1.67 & 2018-04-12 01:57:33 & 0.222 & 2.12\\
4.45 & 174.88 & 1.61 & 2018-04-12 03:53:08 & 0.860 & 2.51\\
1.75 & 175.88 & 1.48 & 2018-04-12 04:36:11 & 0.001 & 2.50\\
2.05 & 174.42 & 1.42 & 2018-04-12 05:19:36 & 0.005 & 1.02\\
\enddata
\tablenotetext{}{\textbf{Note. }The uncertainty corresponds to the 50\% containment. $\Delta \theta$ is the distance from the best-fit HAWC hot-spot position to the measured neutrino position.}
\end{deluxetable*}

We looked at the SIMBAD catalog \citep{simbad} for sources that appear near the coincidences\footnote{For the SIMBAD catalog search, we focus on sources in the 50\% containment region}, and at the Fermi All-sky Variability Analysis (FAVA) online tool\footnote{\url{https://fermi.gsfc.nasa.gov/ssc/data/access/lat/FAVA/}} for any evidence of past flares in the region based on the light curves provided by FAVA. 

For the coincidence of 2016 with FAR of 0.99 per year, there is a radio galaxy in the nearby region, PKS 0017+026 also known as TXS 0017+026~\citep{pks}. This source is 0.04$^{\circ}$ away from the best-fit position of the coincidence. Unfortunately, no distance information is available to estimate the gamma-ray attenuation.  
 Other sources that appear nearby are quasars, but in general these sources are too distant (redshift above 0.3), resulting in strong gamma-ray attenuation.
With the FAVA tool, the source from the 3FGL catalog, J0020.9+0323, was found 0.52$^{\circ}$  away from the best-fit coincidence position, which is outside the 50\% containment region. The 3FGL catalog mentions that this is an unassociated source \citep{3fgl}.

For the coincidence of 2018 with FAR of 0.026 per year, several sources appear in the SIMBAD catalog. There are nine radio galaxies within 0.74 degrees of the best-fit location of the coincidence from the NRAO VLA Sky Survey Catalog. From these, only NVSS J113719+022200 had some information about its distance (redshift of 0.19). We did not find nearby sources in the FAVA monitoring tool for this coincidence.

Both coincidences found with this analysis are therefore consistent with background expectations. Follow-up observations in the optical and X-ray could be helpful to discern if any of these sources are related to the coincident events. 

\subsection{Upper Limit}\label{sec:upperlimit}

Knowing that we observed two coincidences in 3 years of observations, we calculate an upper limit for the parameter space shown in Fig. \ref{fig:sensUL}. We apply Poisson statistics to obtain a 90\% confidence level by using Equation (9.54) in \cite{analysisbook}. This equation gives us an upper limit on the Poisson rate of the signal based on the observation and assuming that in 3 years of observations we expect three coincidences from background.  The  result is a signal Poisson rate $r_S= 3.5$, giving a total Poisson rate of $r_B + r_S = 6.5$. We use the procedure in Sec. \ref{sec:sensitivity} to find the corresponding upper limit values in the parameter space in Fig. \ref{fig:sensUL}. 

\section{Real-time system}\label{sec:system}

The real-time implementation of the analysis started on 2019 November 20. 
As specified in \cite{amonII}, we use the \textit{amonpy} software for the real-time implementation of the analysis. A major difference is that the system is now running at Amazon Web Services (AWS) servers, which will further improve AMON's uptime. We set a threshold for public alerts at an FAR $<4$ coincidences per year. This threshold is set so that there is a reasonable number of statistically interesting coincidences that can be followed up during a year. Alerts are sent immediately to AMON members, and a GCN notice is generated. A GCN circular is also written to inform the rest of the astrophysical community. 
The first public alert of the system was sent out on 2020 February 2. It had an FAR of 1.39 per year. The reported position is (RA, Dec)$=$200.3$^{\circ}$, 12.71$^{\circ}$, with 50\% radius of 0.17$^{\circ}$ \citep[see GCN circular, 26963 ][]{nuem1}. The MASTER Global Robotic Net and the ANTARES observatory performed follow-up observations of the coincidence, but no transient event was observed \citep[see GCN circulars 26973 and 26976][]{nuem2,nuem3}.

The largest latency of the analysis comes from the HAWC analysis of the hot spots, since the transit needs to complete before sending that information to AMON. Based on Fig. \ref{fig:transittime}, the hot-spot duration can last from less than an hour to a bit more than 6 hours. The latency, once the data are in the AMON server, is less than a minute to perform the analysis and send the alert to the public. 

\section{Conclusion}\label{sec:concl}

We developed a method to search for coincidences of subthreshold data from the HAWC and the IceCube observatories. Using coincidences of subthreshold data allows us to recover signal events that cannot be differentiated from the background in each individual detector. The method was tested on archival data taken between the years 2015 and 2018. 
We found two coincidences in the archival analysis that crossed the FAR threshold of one per year, consistent with the background expectations of three coincidences in three years.
Although a few sources were found near the best coincidence positions, these results are still consistent with the expectation from random coincidences. 
The real-time analysis has produced one alert so far, with an FAR of 1.39 per year. It was sent out to the community.
We encourage other observatories to perform follow-up observations of these results and the real-time alerts in the future.

\acknowledgments

\textbf{Acknowledgments}

AMON: This research or portions of this research were conducted with Advanced CyberInfrastructure computational resources provided by the Institute for Computational and Data Sciences at the Pennsylvania State University ( https://www.icds.psu.edu/ ).
This material is based upon work supported by the National Science Foundation under Grants PHY-1708146 and PHY-1806854 and by the Institute for Gravitation and the Cosmos of the Pennsylvania State University. Any opinions, findings, and conclusions or recommendations expressed in this material are those of the author(s) and do not necessarily reflect the views of the National Science Foundation. Felicia Krauss was supported as an Eberly Research Fellow by the Eberly College of Science at the Pennsylvania State University.

HAWC: We acknowledge the support from: the US National Science Foundation (NSF); the US Department of Energy Office of High-Energy Physics; the Laboratory Directed Research and Development (LDRD) program of Los Alamos National Laboratory; Consejo Nacional de Ciencia y Tecnolog\'ia (CONACyT), M\'exico, grants 271051, 232656, 260378, 179588, 254964, 258865, 243290, 132197, A1-S-46288, A1-S-22784, c\'atedras 873, 1563, 341, 323, Red HAWC, M\'exico; DGAPA-UNAM grants IG101320, IN111315, IN111716-3, IN111419, IA102019, IN112218; VIEP-BUAP; PIFI 2012, 2013, PROFOCIE 2014, 2015; the University of Wisconsin Alumni Research Foundation; the Institute of Geophysics, Planetary Physics, and Signatures at Los Alamos National Laboratory; Polish Science Centre grant, DEC-2017/27/B/ST9/02272; Coordinaci\'on de la Investigaci\'on Cient\'ifica de la Universidad Michoacana; Royal Society - Newton Advanced Fellowship 180385; Generalitat Valenciana, grant CIDEGENT/2018/034; Chulalongkorn University’s CUniverse (CUAASC) grant. Thanks to Scott Delay, Luciano D\'iaz and Eduardo Murrieta for technical support.

IceCube: We gratefully acknowledge the following support: USA {\textendash} U.S. National Science Foundation-Office of Polar Programs,
U.S. National Science Foundation-Physics Division,
Wisconsin Alumni Research Foundation,
Center for High Throughput Computing (CHTC) at the University of Wisconsin-Madison,
Open Science Grid (OSG),
Extreme Science and Engineering Discovery Environment (XSEDE),
U.S. Department of Energy-National Energy Research Scientific Computing Center,
Particle astrophysics research computing center at the University of Maryland,
Institute for Cyber-Enabled Research at Michigan State University,
and Astroparticle physics computational facility at Marquette University;
Belgium {\textendash} Funds for Scientific Research (FRS-FNRS and FWO),
FWO Odysseus and Big Science programmes,
and Belgian Federal Science Policy Office (Belspo);
Germany {\textendash} Bundesministerium f{\"u}r Bildung und Forschung (BMBF),
Deutsche Forschungsgemeinschaft (DFG),
Helmholtz Alliance for Astroparticle Physics (HAP),
Initiative and Networking Fund of the Helmholtz Association,
Deutsches Elektronen Synchrotron (DESY),
and High Performance Computing cluster of the RWTH Aachen;
Sweden {\textendash} Swedish Research Council,
Swedish Polar Research Secretariat,
Swedish National Infrastructure for Computing (SNIC),
and Knut and Alice Wallenberg Foundation;
Australia {\textendash} Australian Research Council;
Canada {\textendash} Natural Sciences and Engineering Research Council of Canada,
Calcul Qu{\'e}bec, Compute Ontario, Canada Foundation for Innovation, WestGrid, and Compute Canada;
Denmark {\textendash} Villum Fonden, Danish National Research Foundation (DNRF), Carlsberg Foundation;
New Zealand {\textendash} Marsden Fund;
Japan {\textendash} Japan Society for Promotion of Science (JSPS)
and Institute for Global Prominent Research (IGPR) of Chiba University;
Korea {\textendash} National Research Foundation of Korea (NRF);
Switzerland {\textendash} Swiss National Science Foundation (SNSF);
United Kingdom {\textendash} Department of Physics, University of Oxford.

\vspace{5mm}
\facilities{HAWC, IceCube, AMON}

\software{astropy~\citep{astropy},
        FIRESONG~\citep{firesong},
        numpy~\citep{numpy},
        scipy~\citep{scipy}
        matplotlib~\citep{matplotlib},
        pandas~\citep{pandas},
        amonpy~\citep{amonII}
        }





\interlinepenalty=10000
\bibliography{biblio}

\end{document}

%% file: authors_AMON.tex
\author[0000-0002-2084-5049]{H.A.~Ayala Solares}
\affiliation{Department of Physics, Pennsylvania State University, University Park, PA 16802, USA}

\author{S.~Coutu}
\affiliation{Department of Physics, Pennsylvania State University, University Park, PA 16802, USA}

\author{J.~J. DeLaunay}
\affiliation{Department of Physics, Pennsylvania State University, University Park, PA 16802, USA}

\author{D.~B. Fox}
\affiliation{Department of Physics, Pennsylvania State University, University Park, PA 16802, USA}
\author[0000-0001-8711-1456]{T. Gr{\'e}goire}
\affiliation{Department of Physics, Pennsylvania State University, University Park, PA 16802, USA}

\author[0000-0001-7197-2788]{A.~Keivani}
\affiliation{Department of Physics, Columbia University, New York, NY 10027, USA}
\affiliation{Columbia Astrophysics Laboratory, Columbia University, New York, NY 10027, USA}

\author[0000-0001-6191-1244]{F.~Krau{\ss}}
\affiliation{Department of Physics, Pennsylvania State University, University Park, PA 16802, USA}

\author[0000-0002-7675-4656]{M.~Mostaf\'{a}}
\affiliation{Department of Physics, Pennsylvania State University, University Park, PA 16802, USA}

\author[0000-0002-5358-5642]{K.~Murase}
\affiliation{Department of Physics, Pennsylvania State University, University Park, PA 16802, USA}

\author{C. F.~Turley}
\affiliation{Department of Physics, Pennsylvania State University, University Park, PA 16802, USA}

\collaboration{AMON Team}
\noaffiliation

%% file: authors_HAWC.tex
\author[0000-0003-0197-5646]{A.~Albert}
\affiliation{Physics Division, Los Alamos National Laboratory, Los Alamos, NM, USA }

\author[0000-0001-8749-1647]{R.~Alfaro}
\affiliation{Instituto de F\'{i}sica, Universidad Nacional Aut\'{o}noma de M\'{e}xico, Ciudad de M\'{e}xico, M\'{e}xico }

\author{C.~Alvarez}
\affiliation{Universidad Aut\'{o}noma de Chiapas, Tuxtla Guti\'{e}rrez, Chiapas, M\'{e}xico}

\author{J.R.~Angeles Camacho}
\affiliation{Instituto de F\'{i}sica, Universidad Nacional Aut\'{o}noma de M\'{e}xico, Ciudad de M\'{e}xico, M\'{e}xico }

\author{J.C.~Arteaga-Vel\'{a}zquez}
\affiliation{Universidad Michoacana de San Nicol\'{a}s de Hidalgo, Morelia, M\'{e}xico }

\author{K.P.~Arunbabu}
\affiliation{Instituto de Geof\'{i}sica, Universidad Nacional Aut\'{o}noma de M\'{e}xico, Ciudad de M\'{e}xico, M\'{e}xico }

\author{D.~Avila Rojas}
\affiliation{Instituto de F\'{i}sica, Universidad Nacional Aut\'{o}noma de M\'{e}xico, Ciudad de M\'{e}xico, M\'{e}xico }


\author[0000-0003-3207-105X]{E.~Belmont-Moreno}
\affiliation{Instituto de F\'{i}sica, Universidad Nacional Aut\'{o}noma de M\'{e}xico, Ciudad de M\'{e}xico, M\'{e}xico }

\author[0000-0002-5493-6344]{C.~Brisbois}
\affiliation{Dept. of Physics, University of Maryland, College Park, MD 20742, USA}

\author[0000-0002-4042-3855]{K.S.~Caballero-Mora}
\affiliation{Universidad Aut\'{o}noma de Chiapas, Tuxtla Guti\'{e}rrez, Chiapas, M\'{e}xico}

\author[0000-0002-8553-3302]{A.~Carrami\~{n}ana}
\affiliation{Instituto Nacional de Astrof\'{i}sica, \'{O}ptica y Electr\'{o}nica, Puebla, M\'{e}xico }

\author[0000-0002-6144-9122]{S.~Casanova}
\affiliation{Institute of Nuclear Physics Polish Academy of Sciences, PL-31342 IFJ-PAN, Krakow, Poland }

\author[0000-0002-7607-9582]{U.~Cotti}
\affiliation{Universidad Michoacana de San Nicol\'{a}s de Hidalgo, Morelia, M\'{e}xico }

\author[0000-0001-9643-4134]{E.~De la Fuente}
\affiliation{Departamento de F\'{i}sica, Centro Universitario de Ciencias Exactas e Ingenierias, Universidad de Guadalajara, Guadalajara, M\'{e}xico }

\author{R.~Diaz Hernandez}
\affiliation{Instituto Nacional de Astrof\'{i}sica, \'{O}ptica y Electr\'{o}nica, Puebla, M\'{e}xico }

\author[0000-0001-8451-7450]{B.L.~Dingus}
\affiliation{Physics Division, Los Alamos National Laboratory, Los Alamos, NM, USA }

\author[0000-0002-2987-9691]{M.A.~DuVernois}
\affiliation{Department of Physics, University of Wisconsin-Madison, Madison, WI 53706, USA }

\author{M.~Durocher}
\affiliation{Physics Division, Los Alamos National Laboratory, Los Alamos, NM, USA }

\author[0000-0002-0087-0693]{J.C.~D\'{i}az-V\'{e}lez}
\affiliation{Departamento de F\'{i}sica, Centro Universitario de Ciencias Exactas e Ingenierias, Universidad de Guadalajara, Guadalajara, M\'{e}xico }

\author[0000-0001-7074-1726]{C.~Espinoza}
\affiliation{Instituto de F\'{i}sica, Universidad Nacional Aut\'{o}noma de M\'{e}xico, Ciudad de M\'{e}xico, M\'{e}xico }

\author{K.L.~Fan}
\affiliation{Dept. of Physics, University of Maryland, College Park, MD 20742, USA}

\author[0000-0002-0794-8780]{H.~Fleischhack}
\affiliation{Department of Physics, Michigan Technological University, Houghton, MI, USA }

\author{N.~Fraija}
\affiliation{Instituto de Astronom\'{i}a, Universidad Nacional Aut\'{o}noma de M\'{e}xico, Ciudad de M\'{e}xico, M\'{e}xico }

\author{A.~Galv\'{a}n-G\'{a}mez}
\affiliation{Instituto de Astronom\'{i}a, Universidad Nacional Aut\'{o}noma de M\'{e}xico, Ciudad de M\'{e}xico, M\'{e}xico }

\author{D.~Garcia}
\affiliation{Instituto de F\'{i}sica, Universidad Nacional Aut\'{o}noma de M\'{e}xico, Ciudad de M\'{e}xico, M\'{e}xico }

\author[0000-0002-4188-5584]{J.A.~Garc\'{i}a-Gonz\'{a}lez}
\affiliation{Instituto de Astronom\'{i}a, Universidad Nacional Aut\'{o}noma de M\'{e}xico, Ciudad de M\'{e}xico, M\'{e}xico }

\author[0000-0003-1122-4168]{F.~Garfias}
\affiliation{Instituto de Astronom\'{i}a, Universidad Nacional Aut\'{o}noma de M\'{e}xico, Ciudad de M\'{e}xico, M\'{e}xico }

\author[0000-0002-5209-5641]{M.M.~Gonz\'{a}lez}
\affiliation{Instituto de Astronom\'{i}a, Universidad Nacional Aut\'{o}noma de M\'{e}xico, Ciudad de M\'{e}xico, M\'{e}xico }

\author[0000-0002-9790-1299]{J.A.~Goodman}
\affiliation{Dept. of Physics, University of Maryland, College Park, MD 20742, USA}

\author{J.P.~Harding}
\affiliation{Physics Division, Los Alamos National Laboratory, Los Alamos, NM, USA }

\author{B.~Hona}
\affiliation{Department of Physics and Astronomy, University of Utah, Salt Lake City, UT, USA }

\author[0000-0002-3808-4639]{D.~Huang}
\affiliation{Department of Physics, Michigan Technological University, Houghton, MI, USA }

\author[0000-0002-5527-7141]{F.~Hueyotl-Zahuantitla}
\affiliation{Universidad Aut\'{o}noma de Chiapas, Tuxtla Guti\'{e}rrez, Chiapas, M\'{e}xico}

\author{P.~H{\"u}ntemeyer}
\affiliation{Department of Physics, Michigan Technological University, Houghton, MI, USA }

\author[0000-0001-5811-5167]{A.~Iriarte}
\affiliation{Instituto de Astronom\'{i}a, Universidad Nacional Aut\'{o}noma de M\'{e}xico, Ciudad de M\'{e}xico, M\'{e}xico }

\author[0000-0002-6738-9351]{A.~Jardin-Blicq}
\affiliation{Max-Planck Institute for Nuclear Physics, 69117 Heidelberg, Germany}
\affiliation{Department of Physics, Faculty of Science, Chulalongkorn University, Bangkok 10330, Thailand}
\affiliation{National Astronomical Research Institute of Thailand (Public
Organization), Chiang Mai 50180, Thailand}

\author[0000-0003-4467-3621]{V.~Joshi}
\affiliation{Erlangen Centre for Astroparticle Physics, Friedrich-Alexander-Universit{\"a}t Erlangen-N{\"u}rnberg, D-91058 Erlangen, Germany}

\author[0000-0001-5516-4975]{H.~Le\'{o}n Vargas}
\affiliation{Instituto de F\'{i}sica, Universidad Nacional Aut\'{o}noma de M\'{e}xico, Ciudad de M\'{e}xico, M\'{e}xico }

\author{J.T.~Linnemann}
\affiliation{Dept. of Physics and Astronomy, Michigan State University, East Lansing, MI 48824, USA}

\author[0000-0001-8825-3624]{A.L.~Longinotti}
\affiliation{Instituto Nacional de Astrof\'{i}sica, \'{O}ptica y Electr\'{o}nica, Puebla, M\'{e}xico }

\author[0000-0003-2810-4867]{G.~Luis-Raya}
\affiliation{Universidad Politecnica de Pachuca, Pachuca, Hgo, M\'{e}xico }

\author{J.~Lundeen}
\affiliation{Dept. of Physics and Astronomy, Michigan State University, East Lansing, MI 48824, USA}

\author[0000-0001-8088-400X]{K.~Malone}
\affiliation{Physics Division, Los Alamos National Laboratory, Los Alamos, NM, USA }

\author[0000-0001-9052-856X]{O.~Martinez}
\affiliation{Facultad de Ciencias F\'{i}sico Matemáticas, Benem\'{e}rita Universidad Aut\'{o}noma de Puebla, Puebla, M\'{e}xico }

\author[0000-0001-9035-1290]{I.~Martinez-Castellanos}
\affiliation{Dept. of Physics, University of Maryland, College Park, MD 20742, USA}

\author{J.~Mart\'{i}nez-Castro}
\affiliation{Centro de Investigaci\'on en Computaci\'on, Instituto Polit\'ecnico Nacional, Mexico City, Mexico.}

\author[0000-0002-2610-863X]{J.A.~Matthews}
\affiliation{Dept of Physics and Astronomy, University of New Mexico, Albuquerque, NM, USA }

\author[0000-0002-8390-9011]{P.~Miranda-Romagnoli}
\affiliation{Universidad Aut\'{o}noma del Estado de Hidalgo, Pachuca, M\'{e}xico }

\author[0000-0002-1114-2640]{E.~Moreno}
\affiliation{Facultad de Ciencias F\'{i}sico Matemáticas, Benem\'{e}rita Universidad Aut\'{o}noma de Puebla, Puebla, M\'{e}xico }

\author[0000-0003-1059-8731]{L.~Nellen}
\affiliation{Instituto de Ciencias Nucleares, Universidad Nacional Aut\'{o}noma de M\'{e}xico, Ciudad de M\'{e}xico, M\'{e}xico }

\author[0000-0001-9428-7572]{M.~Newbold}
\affiliation{Department of Physics and Astronomy, University of Utah, Salt Lake City, UT, USA }

\author[0000-0002-6859-3944]{M.U.~Nisa}
\affiliation{Dept. of Physics and Astronomy, Michigan State University, East Lansing, MI 48824, USA}

\author[0000-0001-7099-108X]{R.~Noriega-Papaqui}
\affiliation{Universidad Aut\'{o}noma del Estado de Hidalgo, Pachuca, M\'{e}xico }

\author{A.~Peisker}
\affiliation{Dept. of Physics and Astronomy, Michigan State University, East Lansing, MI 48824, USA}

\author[0000-0001-5998-4938]{E.G.~P\'{e}rez-P\'{e}rez}
\affiliation{Universidad Politecnica de Pachuca, Pachuca, Hgo, M\'{e}xico }

\author[0000-0002-6524-9769]{C.D.~Rho}
\affiliation{Natural Science Research Institute, University of Seoul, Seoul, Republic of Korea}

\author[0000-0003-1327-0838]{D.~Rosa-Gonz\'{a}lez}
\affiliation{Instituto Nacional de Astrof\'{i}sica, \'{O}ptica y Electr\'{o}nica, Puebla, M\'{e}xico }

\author{H.~Salazar}
\affiliation{Facultad de Ciencias F\'{i}sico Matem\'{a}ticas, Benem\'{e}rita Universidad Aut\'{o}noma de Puebla, Puebla, M\'{e}xico }

\author[0000-0002-8610-8703]{F.~Salesa Greus}
\affiliation{Institute of Nuclear Physics Polish Academy of Sciences, PL-31342 IFJ-PAN, Krakow, Poland }
\affiliation{Instituto de F\'{i}sica Corpuscular, CSIC, Universitat de Val\`{e}ncia, E-46980, Paterna, Valencia, Spain}

\author{A.~Sandoval}
\affiliation{Instituto de F\'{i}sica, Universidad Nacional Aut\'{o}noma de M\'{e}xico, Ciudad de M\'{e}xico, M\'{e}xico }

\author{A.J.~Smith}
\affiliation{Dept. of Physics, University of Maryland, College Park, MD 20742, USA}

\author[0000-0002-1492-0380]{R.W.~Springer}
\affiliation{Department of Physics and Astronomy, University of Utah, Salt Lake City, UT, USA }

\author[0000-0001-9725-1479]{K.~Tollefson}
\affiliation{Dept. of Physics and Astronomy, Michigan State University, East Lansing, MI 48824, USA}

\author[0000-0002-1689-3945]{I.~Torres}
\affiliation{Instituto Nacional de Astrof\'{i}sica, \'{O}ptica y Electr\'{o}nica, Puebla, M\'{e}xico }

\author{R.~Torres-Escobedo}
\affiliation{Departamento de F\'{i}sica, Centro Universitario de Ciencias Exactas e Ingenierias, Universidad de Guadalajara, Guadalajara, M\'{e}xico }

\author{F.~Ure\~{n}a-Mena}
\affiliation{Instituto Nacional de Astrof\'{i}sica, \'{O}ptica y Electr\'{o}nica, Puebla, M\'{e}xico }

\author[0000-0001-6876-2800]{L.~Villase\~{n}or}
\affiliation{Facultad de Ciencias F\'{i}sico Matem\'{a}ticas, Benem\'{e}rita Universidad Aut\'{o}noma de Puebla, Puebla, M\'{e}xico }

\author{T.~Weisgarber}
\affiliation{Department of Chemistry and Physics, California University of Pennsylvania, California, Pennsylvania, USA}

\author{E.~Willox}
\affiliation{Dept. of Physics, University of Maryland, College Park, MD 20742, USA}

\author[0000-0001-9976-2387]{A.~Zepeda}
\affiliation{Physics Department, Centro de Investigacion y de Estudios Avanzados del IPN, Mexico City, DF, Mexico }

\author{H.~Zhou}
\affiliation{Tsung-Dao Lee Institute \& School of Physics and Astronomy, Shanghai Jiao Tong University, Shanghai, China}

\author{C.~de Le\'{o}n}
\affiliation{Universidad Michoacana de San Nicol\'{a}s de Hidalgo, Morelia, M\'{e}xico }

\collaboration{HAWC Collaboration}
\noaffiliation

%% file: authors_IC.tex
\affiliation{III. Physikalisches Institut, RWTH Aachen University, D-52056 Aachen, Germany}
\affiliation{Department of Physics, University of Adelaide, Adelaide, 5005, Australia}
\affiliation{Dept. of Physics and Astronomy, University of Alaska Anchorage, 3211 Providence Dr., Anchorage, AK 99508, USA}
\affiliation{Dept. of Physics, University of Texas at Arlington, 502 Yates St., Science Hall Rm 108, Box 19059, Arlington, TX 76019, USA}
\affiliation{CTSPS, Clark-Atlanta University, Atlanta, GA 30314, USA}
\affiliation{School of Physics and Center for Relativistic Astrophysics, Georgia Institute of Technology, Atlanta, GA 30332, USA}
\affiliation{Dept. of Physics, Southern University, Baton Rouge, LA 70813, USA}
\affiliation{Dept. of Physics, University of California, Berkeley, CA 94720, USA}
\affiliation{Lawrence Berkeley National Laboratory, Berkeley, CA 94720, USA}
\affiliation{Institut f{\"u}r Physik, Humboldt-Universit{\"a}t zu Berlin, D-12489 Berlin, Germany}
\affiliation{Fakult{\"a}t f{\"u}r Physik {\&} Astronomie, Ruhr-Universit{\"a}t Bochum, D-44780 Bochum, Germany}
\affiliation{Universit{\'e} Libre de Bruxelles, Science Faculty CP230, B-1050 Brussels, Belgium}
\affiliation{Vrije Universiteit Brussel (VUB), Dienst ELEM, B-1050 Brussels, Belgium}
\affiliation{Dept. of Physics, Massachusetts Institute of Technology, Cambridge, MA 02139, USA}
\affiliation{Dept. of Physics and Institute for Global Prominent Research, Chiba University, Chiba 263-8522, Japan}
\affiliation{Department of Physics, Loyola University Chicago, Chicago, IL 60660, USA}
\affiliation{Dept. of Physics and Astronomy, University of Canterbury, Private Bag 4800, Christchurch, New Zealand}
\affiliation{Dept. of Physics, University of Maryland, College Park, MD 20742, USA}
\affiliation{Dept. of Astronomy, Ohio State University, Columbus, OH 43210, USA}
\affiliation{Dept. of Physics and Center for Cosmology and Astro-Particle Physics, Ohio State University, Columbus, OH 43210, USA}
\affiliation{Niels Bohr Institute, University of Copenhagen, DK-2100 Copenhagen, Denmark}
\affiliation{Dept. of Physics, TU Dortmund University, D-44221 Dortmund, Germany}
\affiliation{Dept. of Physics and Astronomy, Michigan State University, East Lansing, MI 48824, USA}
\affiliation{Dept. of Physics, University of Alberta, Edmonton, Alberta, Canada T6G 2E1}
\affiliation{Erlangen Centre for Astroparticle Physics, Friedrich-Alexander-Universit{\"a}t Erlangen-N{\"u}rnberg, D-91058 Erlangen, Germany}
\affiliation{Physik-department, Technische Universit{\"a}t M{\"u}nchen, D-85748 Garching, Germany}
\affiliation{D{\'e}partement de physique nucl{\'e}aire et corpusculaire, Universit{\'e} de Gen{\`e}ve, CH-1211 Gen{\`e}ve, Switzerland}
\affiliation{Dept. of Physics and Astronomy, University of Gent, B-9000 Gent, Belgium}
\affiliation{Dept. of Physics and Astronomy, University of California, Irvine, CA 92697, USA}
\affiliation{Karlsruhe Institute of Technology, Institut f{\"u}r Kernphysik, D-76021 Karlsruhe, Germany}
\affiliation{Dept. of Physics and Astronomy, University of Kansas, Lawrence, KS 66045, USA}
\affiliation{SNOLAB, 1039 Regional Road 24, Creighton Mine 9, Lively, ON, Canada P3Y 1N2}
\affiliation{Department of Physics and Astronomy, UCLA, Los Angeles, CA 90095, USA}
\affiliation{Department of Physics, Mercer University, Macon, GA 31207-0001, USA}
\affiliation{Dept. of Astronomy, University of Wisconsin, Madison, WI 53706, USA}
\affiliation{Dept. of Physics and Wisconsin IceCube Particle Astrophysics Center, University of Wisconsin, Madison, WI 53706, USA}
\affiliation{Institute of Physics, University of Mainz, Staudinger Weg 7, D-55099 Mainz, Germany}
\affiliation{Department of Physics, Marquette University, Milwaukee, WI, 53201, USA}
\affiliation{Institut f{\"u}r Kernphysik, Westf{\"a}lische Wilhelms-Universit{\"a}t M{\"u}nster, D-48149 M{\"u}nster, Germany}
\affiliation{Bartol Research Institute and Dept. of Physics and Astronomy, University of Delaware, Newark, DE 19716, USA}
\affiliation{Dept. of Physics, Yale University, New Haven, CT 06520, USA}
\affiliation{Dept. of Physics, University of Oxford, Parks Road, Oxford OX1 3PU, UK}
\affiliation{Dept. of Physics, Drexel University, 3141 Chestnut Street, Philadelphia, PA 19104, USA}
\affiliation{Physics Department, South Dakota School of Mines and Technology, Rapid City, SD 57701, USA}
\affiliation{Dept. of Physics, University of Wisconsin, River Falls, WI 54022, USA}
\affiliation{Dept. of Physics and Astronomy, University of Rochester, Rochester, NY 14627, USA}
\affiliation{Oskar Klein Centre and Dept. of Physics, Stockholm University, SE-10691 Stockholm, Sweden}
\affiliation{Dept. of Physics and Astronomy, Stony Brook University, Stony Brook, NY 11794-3800, USA}
\affiliation{Dept. of Physics, Sungkyunkwan University, Suwon 16419, Korea}
\affiliation{Institute of Basic Science, Sungkyunkwan University, Suwon 16419, Korea}
\affiliation{Dept. of Physics and Astronomy, University of Alabama, Tuscaloosa, AL 35487, USA}
\affiliation{Dept. of Astronomy and Astrophysics, Pennsylvania State University, University Park, PA 16802, USA}
\affiliation{Dept. of Physics, Pennsylvania State University, University Park, PA 16802, USA}
\affiliation{Dept. of Physics and Astronomy, Uppsala University, Box 516, S-75120 Uppsala, Sweden}
\affiliation{Dept. of Physics, University of Wuppertal, D-42119 Wuppertal, Germany}
\affiliation{DESY, D-15738 Zeuthen, Germany}

\author{M. G. Aartsen}
\affiliation{Dept. of Physics and Astronomy, University of Canterbury, Private Bag 4800, Christchurch, New Zealand}
\author{R. Abbasi}
\affiliation{Department of Physics, Loyola University Chicago, Chicago, IL 60660, USA}
\author{M. Ackermann}
\affiliation{DESY, D-15738 Zeuthen, Germany}
\author{J. Adams}
\affiliation{Dept. of Physics and Astronomy, University of Canterbury, Private Bag 4800, Christchurch, New Zealand}
\author{J. A. Aguilar}
\affiliation{Universit{\'e} Libre de Bruxelles, Science Faculty CP230, B-1050 Brussels, Belgium}
\author{M. Ahlers}
\affiliation{Niels Bohr Institute, University of Copenhagen, DK-2100 Copenhagen, Denmark}
\author{M. Ahrens}
\affiliation{Oskar Klein Centre and Dept. of Physics, Stockholm University, SE-10691 Stockholm, Sweden}
\author{C. Alispach}
\affiliation{D{\'e}partement de physique nucl{\'e}aire et corpusculaire, Universit{\'e} de Gen{\`e}ve, CH-1211 Gen{\`e}ve, Switzerland}
\author{N. M. Amin}
\affiliation{Bartol Research Institute and Dept. of Physics and Astronomy, University of Delaware, Newark, DE 19716, USA}
\author{K. Andeen}
\affiliation{Department of Physics, Marquette University, Milwaukee, WI, 53201, USA}
% ...

\collaboration{IceCube Collaboration}
\noaffiliation